\documentclass[3p, review, 12pt]{elsarticle}

\title{Secondary Electron Count Imaging in SEM}

\author[add1]{Akshay Agarwal}
\author[add1]{John Simonaitis}
\author[add2]{Vivek K. Goyal}
\author[add1]{Karl K. Berggren}
\address[add1]{Department of Electrical Engineering and Computer Science, Massachusetts Institute of Technology}
\address[add2]{Department of Electrical and Computer Engineering, Boston University}

\usepackage{cmap}
\usepackage[utf8]{inputenc}
\usepackage[T1]{fontenc}
\usepackage{graphicx}
\usepackage{siunitx}
\usepackage{cite}
\usepackage{array}
\usepackage{mathtools}
\usepackage{pgfplots}
\pgfplotsset{compat=1.12}
\usetikzlibrary{arrows}
\usepackage{outlines}
\usepackage{enumitem}
\usepackage{matlab-prettifier}
\usepackage[nottoc]{tocbibind}
\usepackage{setspace}
\usepackage{amssymb}
\usepackage{braket}
\usepackage{dcolumn}
\usepackage[titletoc]{appendix}
\usepackage{float}
\usepackage{bm}
\usepackage{caption}
\usepackage{xr}
\usepackage[normalem]{ulem} 
\makeatletter
\newcommand*{\addFileDependency}[1]{
  \typeout{(#1)}
  \@addtofilelist{#1}
  \IfFileExists{#1}{}{\typeout{No file #1.}}
}
\makeatother

 \usepackage[stable]{footmisc}
\def\all{all}
\ifx\files\all  \else 
\def\um{\si{\micro\meter}}

\def\ms{\si{\milli\second}}
\def\us{\si{\micro\second}}
\def\ns{\si{\nano\second}}

\def\pA{\si{\pico\ampere}}

\def\keV{\si{\kilo\electronvolt}}

\begin{document}


\begin{abstract}
Scanning electron microscopy (SEM) is a versatile technique used to image samples at the nanoscale. Conventional imaging by this technique relies on finding the average intensity of the signal generated on a detector by secondary electrons (SEs) emitted from the sample and is subject to noise due to variations in the voltage signal from the detector. This noise can result in degradation of the SEM image quality for a given imaging dose. SE count imaging, which uses the direct count of SEs detected from the sample instead of the average signal intensity, would overcome this limitation and lead to improvement in SEM image quality. In this paper, we implement an SE count imaging scheme by synchronously outcoupling the detector and beam scan signals from the microscope and using custom code to count detected SEs. We demonstrate a $\sim$30\% increase in the image signal-to-noise-ratio due to SE counting compared to conventional imaging. The only external hardware requirement for this imaging scheme is an oscilloscope fast enough to accurately sample the detector signal for SE counting, making the scheme easily implementable on any SEM\@.
\end{abstract}

\maketitle

\section{Introduction}
\label{sec:intro}
Scanning electron microscopy (SEM) is a powerful imaging modality that is routinely used for nanoscale imaging, analysis, and characterization of a wide variety of samples~\citep{Reimer1998e}. In SEM, an electron beam with energy typically between 0.5 and 30~$\keV$ raster scans over the sample. The position of the electron beam is controlled by electrostatic scan coils, which drive the beam in the horizontal (fast) direction and the vertical (slow) direction. At each scan position (referred to as a pixel), the incident beam causes emission of electrons from the surface of the sample. These secondary electrons (SEs) are detected by an SE detector, which converts the SE intensity from each pixel to a voltage signal. Conventionally, the analog average of this voltage signal is converted to an 8-bit brightness value for each pixel to generate the sample image.

SE count imaging is an alternative SEM imaging scheme in which the analog average of the SE signal intensity would be replaced by the direct count of SEs per pixel. Such an imaging scheme has the potential to provide higher image quality than conventional SE imaging for a given dose of incident electrons~\citep{Joy2007}. Counting SEs from each pixel would eliminate noise in the image arising from variations in the voltage signal generated by the SE detector and subsequent processing steps. Therefore, an SE count imaging scheme would be beneficial for all types of SE imaging, particularly the imaging of radiation-sensitive samples such as proteins and biomolecules where the imaging dose restricts the achievable image quality~\citep{Egerton2004, Egerton2019}. 

SE count imaging in SEMs was pioneered by Yamada and co-workers~\citep{Yamada1990, Yamada1990a, Yamada1991, Yamada1991a, Uchikawa1992}. In their work, the SE detector signal was coupled to external discriminator and pulse-counter circuits to count the number of SEs for each scan position on the sample. The collection, processing, and readout of pulses in this circuit was synchronized with the SEM scan position at each pixel. This setup was used to generate SE count images of different types of organic and inorganic samples and to demonstrate an improvement in signal-to-noise-ratio (SNR) for digital imaging compared to conventional imaging. However, this implementation did not lead to SE counting being incorporated into commercial SEM imaging. An important factor responsible for the unavailability of SE count imaging was the external circuitry required for counting SEs and synchronizing the counting with the SEM scan coils. Developing these external circuits and making them compatible with different SEM hardware and software can be a challenge. Furthermore, the requirement for synchronizing electronics also limited this experiment to long pixel dwell times (typically tens of $\us$). More recently, SE image histograms have been investigated for counting SEs~\citep{Joy2007, Agarwal2021}, motivated in part by similar work in scanning transmission electron microscopy (STEM)~\citep{Ishikawa2014,Sang2016, Mittelberger2018, Mullarkey2020}. However, such techniques are limited by the same voltage signal variations that are present in conventional SE imaging. Therefore, an accurate SE counting method that does not require external circuits and nanosecond-scale synchronization is of interest and would enable more widespread adoption of SE count imaging.

In this work, we will demonstrate an SE count imaging scheme by synchronously outcoupling the SE detector and SEM beam scan signals onto an oscilloscope, and processing the outcoupled data using custom MATLAB code to count the number of SEs detected for each sample pixel. Using this technique, we imaged a copper mesh sample and demonstrated a $\sim$30\% increase in the image signal-to-noise-ratio for SE count imaging compared to conventional SEM imaging. The only external hardware requirement to implement this scheme was an oscilloscope; we did not use any $\ns$-scale synchronization circuits unlike previous SE count imaging schemes. Therefore, this SE count imaging scheme could potentially be implemented on many different SEM systems and incorporated with standard SEM software.

\section{Experimental Methods}
\label{sec:expmeth}
In this section we will describe the experimental setup as well as the processing steps we took to obtain SE count images. We used a Zeiss LEO 1525 SEM for all the results reported in this paper. We will refer to the SEM scan voltage that scans the beam horizontally across the sample as the \textit{horizontal linescan waveform}. An image is computed from the SE detector signal recorded for a series of such horizontal linescan waveforms. We will refer to the horizontal linescan waveform and the SE detector signal collected during one scan over the sample area being imaged as an \textit{acquisition frame}.   
\begin{figure*}
    \centering
    \includegraphics[scale=0.48]{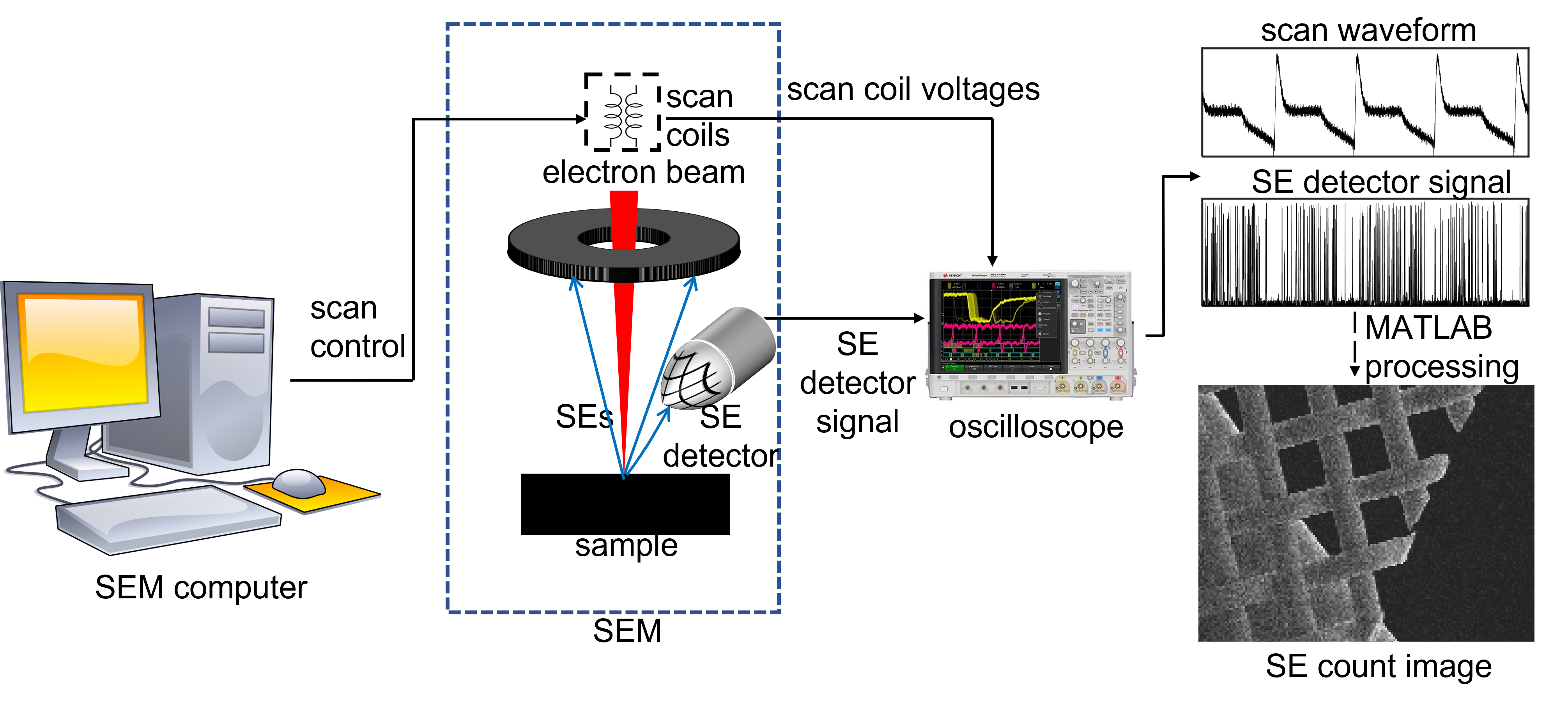}
    \caption{Experimental setup for SE count imaging. The SEM computer controls the scan area and speed on the SEM\@. We outcoupled both the scan coil voltages and the SE detector signal onto a 2~GHz oscilloscope. Data collection on the oscilloscope was triggered by the sawtooth-shaped scan coil voltage signal. Each acquisition frame consisted of the scan waveform and the SE detector signal with their time axes referenced to the same retrace spike. We collected several such acquisition frames and processed them using custom MATLAB code to generate the final SE count image.}
    \label{fig:fig1}
\end{figure*}

\subsection{Imaging setup}
Figure~\ref{fig:fig1} is a schematic of our SE count imaging setup. As is usually done in SEM, we used the SEM computer to control beam scan parameters such as scan speed and scan area (\textit{i.e.}, magnification). The SEM we used was equipped with two SE detectors -- an in-chamber detector and an in-lens detector. We outcoupled the signal from both SE detectors as well as the SEM scan coil voltage waveforms to three channels of an oscilloscope (2~GHz LeCroy WaveRunner 6200A) using BNC cables. This outcoupling allowed us to read out full acquisition frames from the SEM\@. The linescan waveform consists of three regions: (1) the trace, during which the voltage is ramped and the electron beam scans across the sample, (2) the retrace spike, when the electron beam is moved back to the start of the line, and (3) a short rest period before the beam starts scanning again. We discuss features of the scan waveform in more detail in Section~S1 of the Supplementary Information. 

We configured the collection of an acquisition frame on the oscilloscope to be triggered by the retrace spike of the first horizontal linescan voltage waveform in the frame. This trigger voltage setting provided a common time axis for both the linescan waveform and the SE detector signal within an acquisition frame. There was some variation in the exact time at which signal collection was triggered due to noise in the linescan signal, which led to misalignment between successive acquisition frames. In Section~\ref{subsec:framealign} we will discuss how this variation affected the data we collected and how we corrected for it in the MATLAB code. 

We specified the data collection time for an acquisition frame on the oscilloscope to be a few $\ms$ longer than the total scan time for the area we were imaging to ensure that the entire scan area was contained in each acquisition frame. The typical acquisition frames we collected had a pixel resolution of around $200\times200$ pixels, and we used a pixel dwell time of 440~$\ns$ for all data acquisition. Therefore, the data collection time for each acquisition frame was on the order of 20~\ms. As discussed in Section FIXME, we found an oscilloscope sampling time of 10~$\ns$~to be adequate for sampling the SE detector signal pulses. Therefore, any oscilloscope with a sampling frequency over 100 MHz could be used to collect acquisition frames. Over a total collection time of 20~\ms, the sampling rate of 10~$\ns$~corresponded to $2\times10^6$ signal samples. This number is close to the maximum number of samples that could be stored on our oscilloscope at a time. This limitation in the maximum number of samples per acquisition frame limited the pixel resolution of the images we could capture. 

We collected 32 acquisition frames from the same scan area on both the in-chamber and in-lens SE detectors to construct SE count images of that area, at a beam current of 2.3~$\pA$ and a pixel dwell time of 440~\ns. We had found the detective quantum efficiency (DQE)~\citep{Oatley1985, Joy2007} of the in-chamber detector to be 0.16 and that of the in-lens detector to be 0.32 previously~\citep{Agarwal2021}, for the working distance used in this work. To generate SE count images, we added the SE counts obtained from these two detectors. We note that saving the data corresponding to each acquisition frame on the oscilloscope took several seconds. We manually blanked the incident electron beam during this time to minimize sample exposure between acquisition frames. 

As discussed earlier, the acquisition frames we collected contained the SEM linescan as well as the SE detector signal referenced to a common time axis. In the rest of this section, we will present an analysis of the SE detector signal to establish that it can be used to count SEs, and then discuss the MATLAB code we used to generate SE count images from acquisition frames. Section~\ref{subsec:detector_signal} shows an analysis of the SE detector signal. In the MATLAB code we addressed several challenges in order to generate SE count images. First, we needed to account for slight variations in the triggering time on the oscilloscope due to noise in the linescan waveform, which we will describe in Section~\ref{subsec:framealign}. Second, we needed to extract the section of the SE detector signal that corresponded to the trace section of the linescan where the incident beam was actually scanning over the sample. We will describe this extraction in Section~\ref{subsec:usefulsignal}. With these challenges addressed, we counted the number of SE pulses for every pixel in each acquisition frame. We will describe our counting algorithm in Section~\ref{subsec:countingpulses}. 

\subsection{Analysis of SE detector signal}
\label{subsec:detector_signal}

\begin{figure*}[ht]
    \centering
    \includegraphics[scale=0.48]{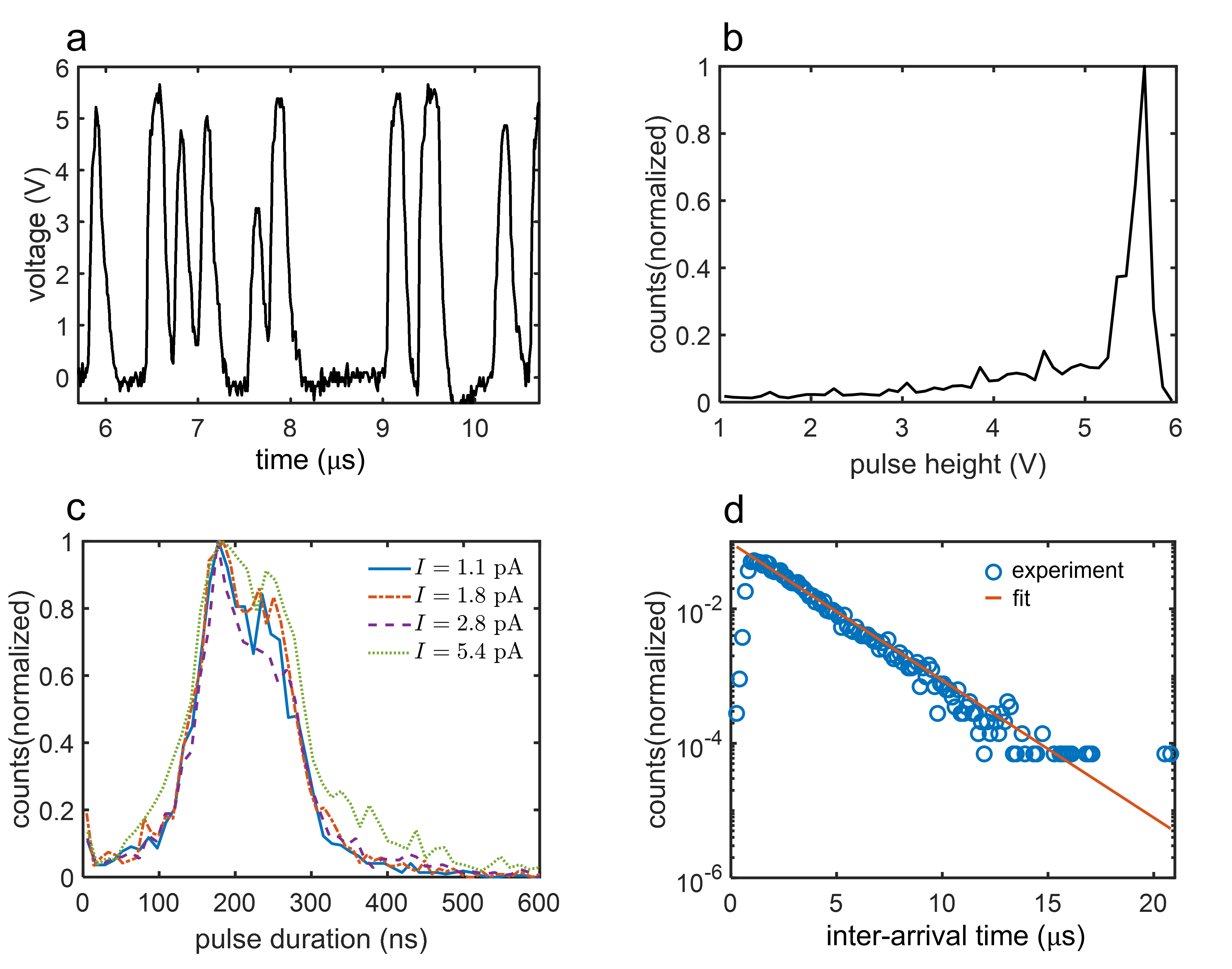}
    \caption{Statistics of SE detector signal. (a) $5$ \si{\micro\second} output signal from the in-chamber detector on the oscilloscope (sampled at 10~\ns) showing pulses due to detected SEs. (b) Histogram of signal pulse heights showing that most pulses are saturated at 5.6~V\@. (c) Histograms of the FWHM pulse durations for beam current $I$ between 1.1~$\pA$ and 5.4~\pA. All histograms have a mean of 180~\ns. (d) Distribution of inter-arrival times of SE pulses on the in-chamber detector. Measured values are displayed as blue unfilled circles. The orange line is an exponential fit with a parameter of 0.47/$\us$. This fit parameter is close to the expected SE count rate of 0.45~SEs/$\us$, as discussed in the text.}
    \label{fig:fig2}
\end{figure*}

Figure \ref{fig:fig2}(a) shows a 5~$\us$~snapshot of the in-chamber SE detector signal, showing voltage pulses from incident SEs. We collected this data at an incident beam current of 2.3~$\pA$ and a pixel dwell time of 440~$\ns$, scanning over a homogeneous sample made of aluminum. In previous work, we had found that an imaging current of 2.3~$\pA$ was well within the linear response region of our SE detectors and had also described the statistics of these voltage pulses~\citep{Agarwal2021}. We had concluded that each pulse corresponds to one detected SE, thereby allowing the counting of SEs. Here, we reiterate the measurements which led to this conclusion. 

At the beam current of 2.3~$\pA$ used in our experiments, assuming that the sample SE yield is 0.2~\citep{Seiler1983} and the detector DQE is 0.16, the mean rate of SEs incident on the detector should be 0.45/$\us$. An analysis of the image histogram statistics confirms this mean SE rate, as described in Section~S2 of the Supplementary Information. Therefore, assuming that the probability distribution function for the number of SEs detected per pixel is roughly Poissonian, the probability of getting multiple SEs during one pixel dwell time is $\sim 1.7\%$~\citep{Frank2005, Sakakibara2019}. Furthermore, the low SE yield results in the probability of one incident beam electron emitting multiple SEs being low (<2\%), further indicating that each detected voltage pulse comes from at most one SE\@.

Analysis of the width statistics of these voltage pulses presented more evidence for our conclusion. The image brightness and contrast settings we used resulted in most pulses having a maximum voltage around 5.6~V, as can be seen in the pulse height distribution plotted in Figure \ref{fig:fig2}(b). This saturated height distribution was advantageous while counting pulses, as described in Section~\ref{subsec:countingpulses}. Figure \ref{fig:fig2}(c) shows the distribution of the full-width-at-half-maximum (FWHM) width of these pulses at different values of the beam current from 1.1~$\pA$ to 5.4~$\pA$. We observed the pulse width distribution remained unchanged as the current increased, indicating that each pulse corresponds to one SE\@. The mean FWHM pulse width was about 180~$\ns$. Based on this mean FWHM pulse width, we used an oscilloscope sampling time of 10~$\ns$ for all acquisition frames in our experiments.

Further evidence is provided in Figure \ref{fig:fig2}(d), which shows the distribution of pulse inter-arrival times. These inter-arrival times were measured as the time duration between the instants when two successive pulses reached half their maximum values. The inter-arrival times for a process following a Poisson distribution with mean $\lambda$ are expected to obey an  exponential distribution given by $\lambda e^{-\lambda t}$. As can be seen in the figure, the inter-arrival time distribution is indeed well-fitted by an exponential distribution for inter-arrival times greater that $\sim $1 $\us$. The decay rate $\lambda$ for this exponential distribution is 0.47/$\us$. This rate is very close to the expected SE detection rate calculated earlier (0.45/$\us$), which again shows that each voltage pulse corresponds to one detected SE\@. We believe that the deviation between the observed pulse inter-arrival times distribution and the exponential fit below 1~$\us$ was a result of limitations in the speed of the scintillator and signal processing electronics in the SE detector.

Having described the evidence for the each of the  voltage pulses arising from individual SEs, we will now describe features of our MATLAB code to generate SE count images from the collected acquisition frames.

\subsection{Aligning acquisition frames and finding linescan period}
\label{subsec:framealign}
\begin{figure*}
    \centering
    \includegraphics[scale=0.48]{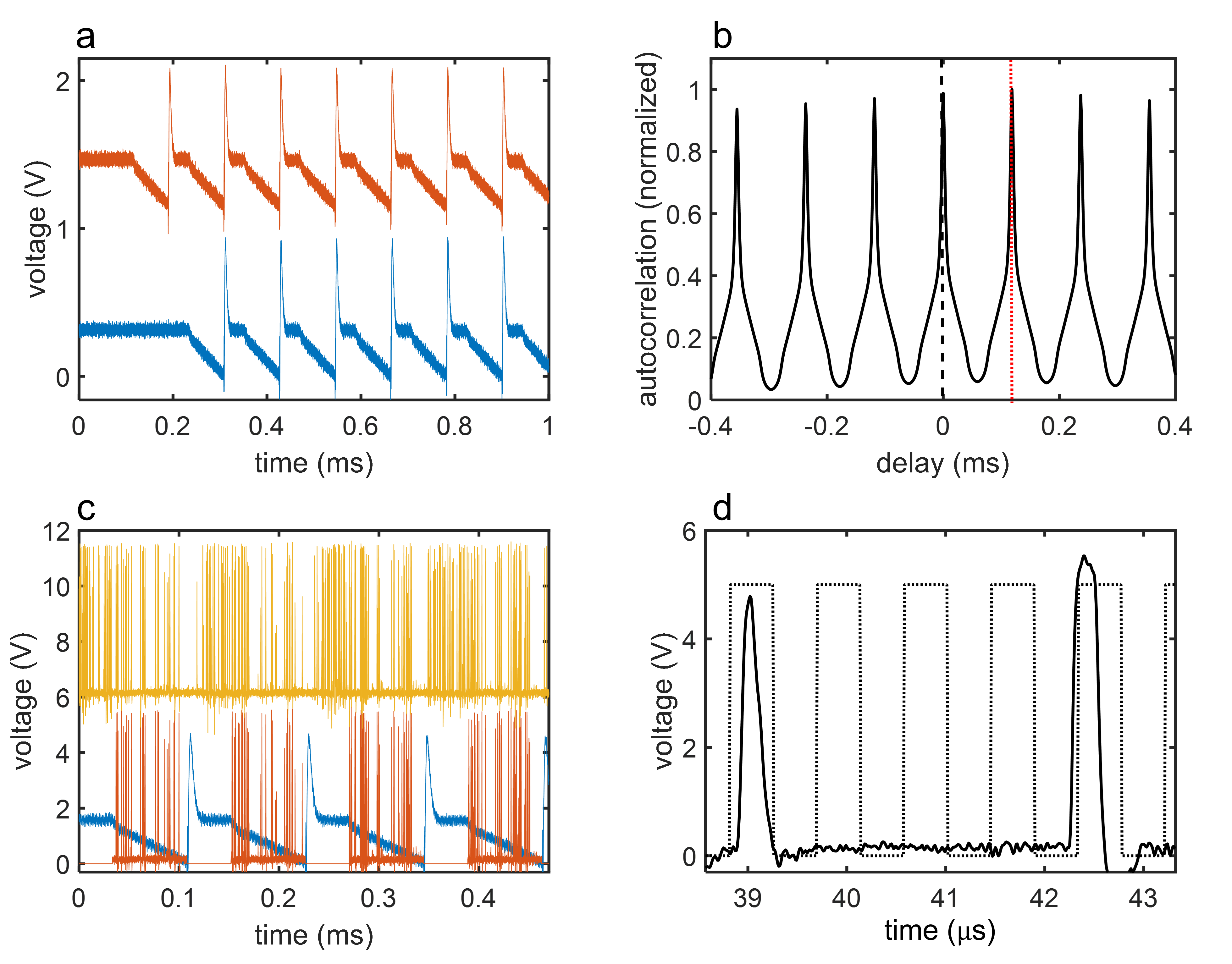}
    \caption[Acquisition frame alignment and SE pulse extraction]{Acquisition frame alignment and SE pulse extraction. (a) Linescans from two acquisition frames at a pixel dwell time of 440~\ns. The top linescan (orange) is misaligned with the bottom linescan (blue) by one line. We added a vertical offset of 1.1~V to the orange linescan for ease of viewing. (b) Cross-correlation of the two linescans in (a). The offset between zero delay (indicated by the dashed black line) and the highest cross-correlation peak (indicated by the dotted red line) equals the misalignment of 118.4~$\us$ between the two linescans. (c) Extraction of useful SE signal, \textit{i.e.}, the sections of the raw SE signal (yellow) corresponding to the trace section of the linescans, during which the incident beam scans over the sample (orange). We added a vertical offset of 6~V to the raw SE detector signal for ease of viewing. (d) Counting SE pulses. The dotted black square waveform shows successive image pixels, with even pixels assigned a value of 0 and odd pixels a value of 5\@. The first of the SE pulses (solid black) is within an odd pixel. However, although the second pulse originates in an even pixel, most of its intensity is present in the next (odd) pixel.}
    \label{fig:fig3}
\end{figure*}

A major challenge we addressed in the MATLAB code for SE count imaging was temporal alignment of all the acquisition frames. Figure~\ref{fig:fig3}(a) shows the linescan waveforms for the first few lines for the first (orange curve) and second (blue curve) acquisition frames captured on the oscilloscope during one imaging experiment. Note that in this figure we added a vertical voltage offset to the scan waveform for the first acquisition frame for ease of viewing. We can see that the two frames were misaligned by a time duration corresponding to one linescan. As we had discussed in Section~\ref{sec:expmeth}, we set the trigger level on the oscilloscope to be close to the peak voltage of the spike that followed each linescan (this spike can be seen at the end of each of the voltage ramps, which correspond to the horizontal scan, in Figure~\ref{fig:fig3}. The misalignment in Figure~\ref{fig:fig3}(a) was caused by noise in the linescan waveform. This noise caused the exact trigger level to be first reached on different spikes for different frames. Each acquisition frame still had SE detector signal and scan waveform referenced to the same time axis, but the misalignment caused the time axis to differ for different frames. We ensured that this misalignment between frames was small enough that the signal from the whole image frame was still captured in the acquisition frame on the oscilloscope.

In the SE count imaging code, we decided to use the linescan waveform for the first frame as the absolute reference to measure the misalignment of all other frames. We used the one-dimensional cross-correlation between the linescans for first frame and those for each of the succeeding frames to measure the misalignment between that frame and the first frame. Figure~\ref{fig:fig3}(b) is the cross-correlation between the linescans for the two acquisition frames from Figure~\ref{fig:fig3}(a). The cross-correlation of two input signals is the product of the two signals as a function of a delay introduced in one of the signals. As the delay changes, one signal `slides over' the other, and the cross-correlation magnitude depends on how similar the two signals are at that delay. Since the two linescan waveforms were periodic, the cross-correlation showed periodic peaks. Figure~\ref{fig:fig3}(b) shows a few of these peaks around zero delay. The highest of these peaks occurred for a delay at which the two linescans were exactly aligned with each other. In Figure~\ref{fig:fig3}(b), we have indicated zero delay with a dashed black vertical line. The delay for the highest peak was 118.4~\us, indicated by the dotted red vertical line. The lower bound on the alignment precision of this technique was equal to the oscilloscope sampling time of 10~\ns. Using this technique we extracted the misalignment for the linescan in each acquisition frame and delayed or advanced the scan waveform and the SE detector signal for that frame by this misalignment to ensure that all frames were aligned to the same time axis.

An additional advantage of calculating the cross-correlation was that the gap between successive peaks of the autocorrelation gave us the periodicity of the linescan waveform. We averaged the values of the gap between the 20 highest cross-correlation peaks to get the value of the linescan period. This period would have been much more difficult to extract from the waveform directly due to noise on the signal, and the cross-correlation was a much more accurate way to measure the periodicity. For the linescans shown in Figure~\ref{fig:fig3}(a), we measured this period to be 118.4~\us. Knowing the pixel dwell time, we also extracted the image resolution from this period. We verified the extracted image resolution by directly analyzing the linescan waveform, as described in Section~S3 of the Supplementary Information. 

\subsection{Finding linescan duration and extracting SE signal during scanning}
\label{subsec:usefulsignal}
The final step before counting the number of SE pulses for each image pixel was determining the start and end time of the trace section of each linescan in every acquisition frame. The trace section of the linescan corresponds to the part of the linescan voltage waveform where the voltage is ramped to scan the beam across the sample. This step was important because in addition to the expected signal pulses during the trace section of the linescan, the SE detector also recorded signal pulses during the other sections of the linescan waveform. Figure~\ref{fig:fig3}(c) shows the raw signal from the in-chamber SE detector (in yellow) and the scan waveform (in blue). Note that in this figure we added an offset of 6~V to the raw SE signal for ease of viewing. We can see that there are signal pulses during the entire waveform, and not just in the trace (\textit{i.e.} ramp) region of the waveform. These signal pulses originated from the sample material at the rest position of the incident electron beam before it started the next linescan. Since this signal does not correspond to the sample region being imaged, it should not be used to generate the image. By finding the start and end time of the trace sections in all the linescans, we can exclude this signal from our analysis and count only the pulses that were recorded when the beam was scanning. Furthermore, having determined the number of pixels per line as described in the previous section, we can use these start and end times as references to segment the signal from one linescan into pixels.

We found the start and end time of the trace section of first linescan in the first acquisition frame manually. With all the acquisition frames aligned and knowing the period of the linescan waveform (as discussed in the previous section), the MATLAB code automatically determined the start and end times of the trace sections in all other linescans in all frames. Once we had determined these times we could determine which sections of the SE signal were acquired during the trace sections and ignore the rest of the detector signal. Figure~\ref{fig:fig3}(c) shows the sections of the in-chamber detector signal acquired during the trace section in red (without any offset). These sections line up with the trace section of the linescan waveform.  After extracting these sections, we divided the trace period into pixels using the extracted number of pixels as described in the previous section. With this segmentation done, our code was ready to count the number of pulses in each pixel.

\subsection{Counting SE pulses}
\label{subsec:countingpulses}

We  generated SE count images by counting the number of SE pulses in each pixel in our code. As described in Section~\ref{subsec:detector_signal}, we used contrast and brightness settings so that most of the SE pulses were saturated at a voltage of 5.6~V\@. This saturation enabled us to use a simple threshold in our code to filter out low voltage noise pulses and count SE pulses originating from the sample. In the MATLAB code we set this threshold voltage to be 1~V\@.

Figure~\ref{fig:fig3}(d) shows an example of the in-chamber detector signal pulses (solid black). In this figure, we have also plotted a square waveform corresponding to successive pixels that the code segmented the trace section of the linescan into (dotted black curve), with odd pixel numbers arbitrarily assigned a value of 5 and even pixels a value of 0 for ease of viewing. We can see that the first SE detector pulse originates in one of the odd pixels and is fully contained within it. The SE imaging code counted one SE in that pixel for this frame. There were no SE pulses for the next 6 pixels. The next SE pulse originated close to the end of an even pixel, but most of its intensity was present in the next (odd) pixel. We assigned this pulse to the even pixel it originated in. The conventional imaging scheme used by the SEM computer would have accounted most of its intensity in the next, odd pixel, leading to inaccuracy in the displayed intensity of that pixel. This `spillover' effect has been discussed in the context of STEM imaging~\citep{Mittelberger2018} and is one of the reasons we expect SE count images to be more accurate that conventional SEM images.

We found the number of SE counts per pixel for each of the 32 acquisition frames for both the in-lens and in-chamber detectors, and added the counts to obtain the final image. We note that  SEMs typically allow the user to ``mix'' the signal from the two detectors in a desired ratio; our method of adding the counts corresponds to a 50/50 mixing.

\section{Results}
\begin{figure*}
    \centering
    \includegraphics[scale=0.48]{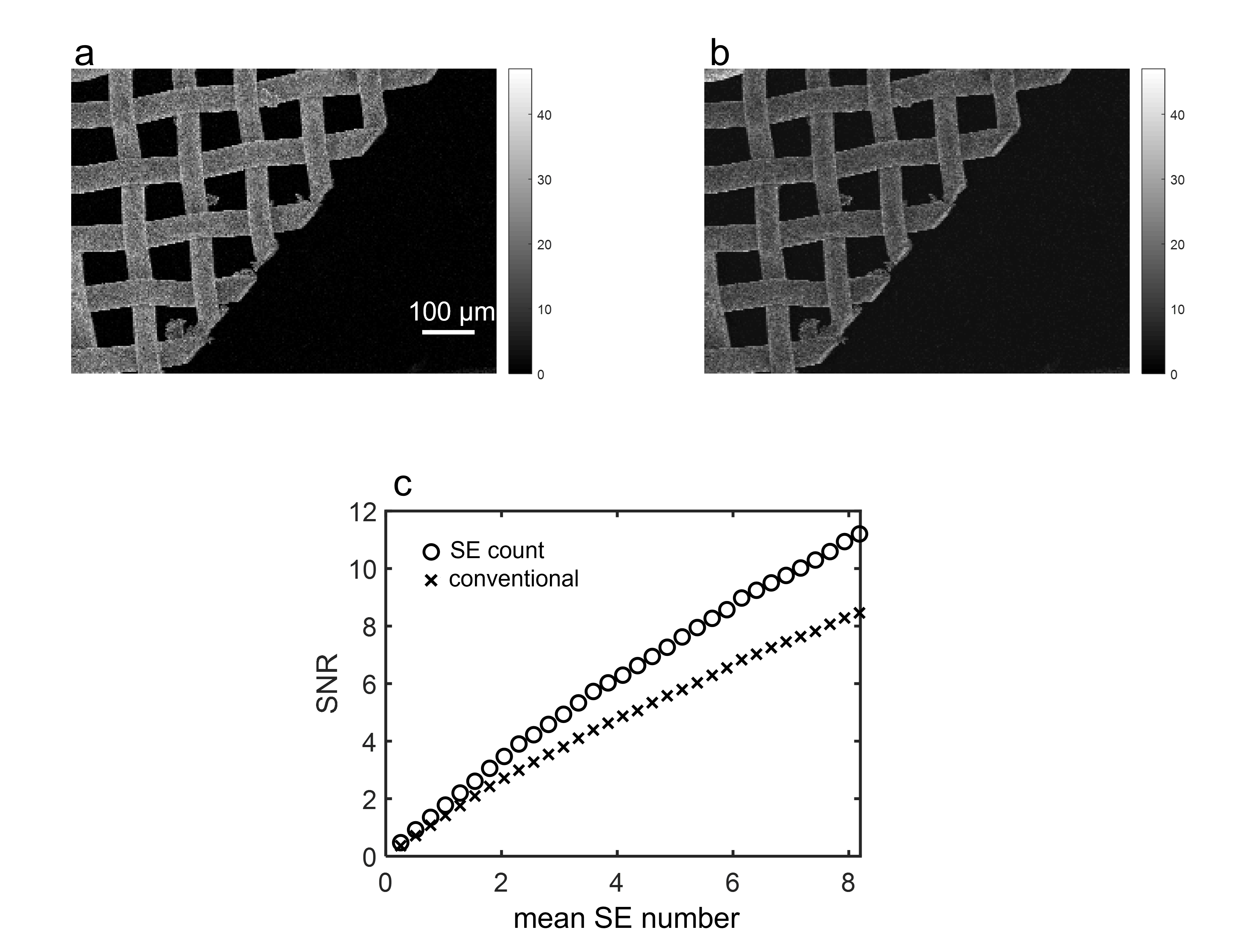}
    \caption{SE count imaging. (a) SE count image of a 120~$\um$ period copper mesh suspended over vacuum.  We generated this image by adding the SE counts from the in-chamber and in-lens detectors over all 32 acquisition frames. The frames were collected at an incident beam current of 2.3~\pA, energy of 10~\keV, and a pixel dwell time of 440~\ns. The image pixel resolution is $262\times188$. The mean SE count in the image is 8.18\@. (b) Conventional image of the same copper mesh grating under the same imaging conditions and scaled to the same mean pixel intensity as the mean SE count in (a). This image shows lower contrast between the copper mesh and background than (a). (c) SNR (refer text for definition) as a function of the mean SE number for the SE count (unfilled circles) and conventional (crosses) images. The SNR increases with mean SE number for both images and is always $\sim 30\%$ higher for the SE count image. The SNR for the SE count image is also more linear than the SNR for the conventional image.}
    \label{fig:fig4}
\end{figure*}

To demonstrate our SE count imaging technique, we chose a sample consisting of a freestanding copper mesh with a period of 120~$\um$ suspended over vacuum. We chose this sample so that we could compare the quality of the SE count image with the conventional image based on the contrast between the image pixels representing the copper mesh and those representing the background vacuum, as well as easily evaluate the SNR of the image. We used an imaging current of 2.3~$\pA$ and a pixel dwell time of 440~$\ns$. 

Figure~\ref{fig:fig4}(a) is an SE count image of the copper mesh generated using the combined SE signal on the in-chamber and in-lens detectors summed over 32 acquisition frames. The pixel resolution of this image is $262\times188$. The colorbar in Figure~\ref{fig:fig4}(a) indicates the number of SEs for a given pixel greyscale level. The maximum number of SEs for any pixel in the image was 50 (summed over all 32 acquisition frames), and the mean number of SEs per pixel in the image was 5.05\@. 

Figure~\ref{fig:fig4}(b) is an image of the same sample generated by finding the average signal for every pixel using the same acquisition frame dataset as Figure~\ref{fig:fig4}(a). To obtain this image, we found the average signal level for every pixel across all 32 frames. This image represents a conventional SEM image; we checked that the statistics of this image were close to those of an image of this sample generated by the SEM software. We scaled the pixel intensities in this image so that the mean intensity in this image was equal to the mean SE count in Figure~\ref{fig:fig4}(a), \textit{i.e.,} 5.05\@. 

From Figures~\ref{fig:fig4}(a) and (b), we can see that the contrast of the copper mesh compared to the background vacuum appears to be lower in the SE count image compared to the conventional image. We numerically evaluated the contrast by dividing the sample into two types of pixels -- `sample` pixels representing the copper mesh and `background` pixels representing the background vacuum. Next, we found the mean pixel intensity for the sample ($I_{\textrm{sample}}$ and the background $I_{\textrm{background}}$ for both the SE count and conventional image (for the SE count image, this mean intensity corresponds to the mean number of detected SEs). Finally, we evaluated the contrast $K$ as~\citep{Boreman2001}:
\begin{equation*}
    K = \frac{I_{\textrm{sample}}-I_{\textrm{background}}}{I_{\textrm{sample}}+I_{\textrm{background}}}
\end{equation*}
Using this technique, we obtained a contrast of 0.64 for the conventional image and 0.95 for the SE count image, demonstrating the superior image quality for the counting technique.

As a second metric to compare the quality of the SE count and conventional images, we evaluated their SNR using a technique first developed by Thong et al.~\citep{Thong2001}. We estimated the signal and noise contributions in the SE image by considering the image autocorrelation function. By varying the number of acquisition frames used to compute the final image, we obtained SNR values for both SE count and conventional imaging for different mean SE counts. We describe this method in more detail in Section~S4 of the Supplementary Information.

Figure~\ref{fig:fig4}(c) is a plot of the SNR for the SE count (unfilled circles) and conventional image (crosses) as a function of the mean SE number. This mean SE number is equal to the mean SE count for the SE count image, and the mean pixel intensity for the conventional image. We can see that for both images the SNR increases as a function of the mean SE number. Furthermore, the SNR for the SE count image is always $\sim30\%$ higher than the SNR for conventional imaging. For example, an SNR of 8 is first achieved for SE count imaging at a mean SE count of 5.38, while for conventional imaging it is achieved at an SE count of 7.68\@. Since the mean SE count scales linearly with the incident beam current~\citep{Agarwal2021}, this difference represents an incident electron dose reduction of 30\% due to SE count imaging for the same SNR\@. 

\section{Conclusions}

In this paper, we implemented SE counting using a 32-acquisition-frame dataset that we acquired by synchronizing the collection of SE detector signal and the SEM scan function on an oscilloscope. We developed code to process this raw dataset to generate SE count images. Our implementation of SE count imaging increased image contrast by 48\% and SNR by 30\% compared to conventional imaging. The major advantage of previous SE counting schemes is that we captured all the fast timing and processing complexity in the SE count imaging code instead of needing to implement it in hardware, which makes our scheme easy to implement on any SEM\@. Extending our SE scheme to larger acquisition frame sizes, as well as live implementation of the scheme will be the subject of future work. Potential designs for live SE count imaging schemes are discussed in Section~S6 of the Supplementary Information.

Our implementation of SE count imaging relies on the observation that each pulse in the SE detector voltage waveform corresponds to one detected SE\@. This observation allowed us to use imaging conditions that saturated the heights of most voltage pulses, and count the number of voltage pulses per pixel to obtain a count of the number of SEs emitted from that pixel. This observation is critically dependent on the low imaging current (2.3~$\pA$) and sample SE yield (assumed to be 0.2~\citep{Seiler1983}) in our experiments, as we have already discussed in Section~\ref{subsec:detector_signal}. In other charged-particle imaging techniques such as helium-ion microscopy (HIM), this observation would no longer be valid, due to the much higher SE yield of such modalities~\citep{Notte2006, Yamanaka2009}. Each voltage pulse could then correspond to multiple SEs. In this case, correlations between the height of the voltage pulse and the number of SEs that produce it could be used to count SEs, provided that the scintillator and amplifier used in the detector is sufficiently linear.

In addition to imaging, an immediate application of our SE count method is the study of deviations from Poisson behavior in the statistics of SEs~\citep{Kurrelmeyer1937, Everhart1959, Oatley1981,Oatley1985, Baumann1980,Baumann1981, Novak2009, Frank2005, Sakakibara2019}. The study of such deviations is important for theoretical modelling of SE emission and imaging. In Section~S5 of the Supplementary Information, we present a preliminary analysis of the probability distribution of SE counts obtained from the experiments reported in this paper, and extracted the degree of deviation from Poisson statistics. We found that the variance of the SE count distribution for the pixels containing the copper mesh was larger than the mean SE counts by a factor of 1.4, indicating deviations from ideal Poisson statistics. A related application of our method is the measurement of the sample SE yield, if the DQE of the SE detector has been characterized previously. By imaging a uniform sample of the material and extracting the mean SE counts using our technique, the SE yield can be directly measured.

\section*{Acknowledgements}
The authors thank Navid Abedzadeh for help with the oscilloscope measurements, and the QEM-2 collaboration for insightful discussions. The authors also acknowledge Minxu Peng, James LeBeau, Dirk Englund, Tony Zhou, Marco Turchetti, and Phillip Keathley for helpful feedback. This work was supported by the Gordon and Betty Moore Foundation. This material is based upon work supported by the U.S. National Science Foundation under Grant No.~181596 and a Graduate Research Fellowship under Grant No.~1745302.
\bibliography{references}
\bibliographystyle{unsrt}

\end{document}


\maketitle

\section{Analysis of SEM scan waveforms}
\label{subsec:scanfeat}
As discussed in the paper, analyzing the different regions of the SEM linescan waveform was a crucial step in our imaging technique, since it allowed us to properly synchronize acquisition frames, extract the resolution of the images, and count only those SEs that were detected when the beam was scanning. In this section, we will discuss the properties of the SEM scan waveforms in more detail. 

\begin{figure*}
    \centering
    \captionsetup{labelformat=empty}
    \includegraphics[scale=0.48]{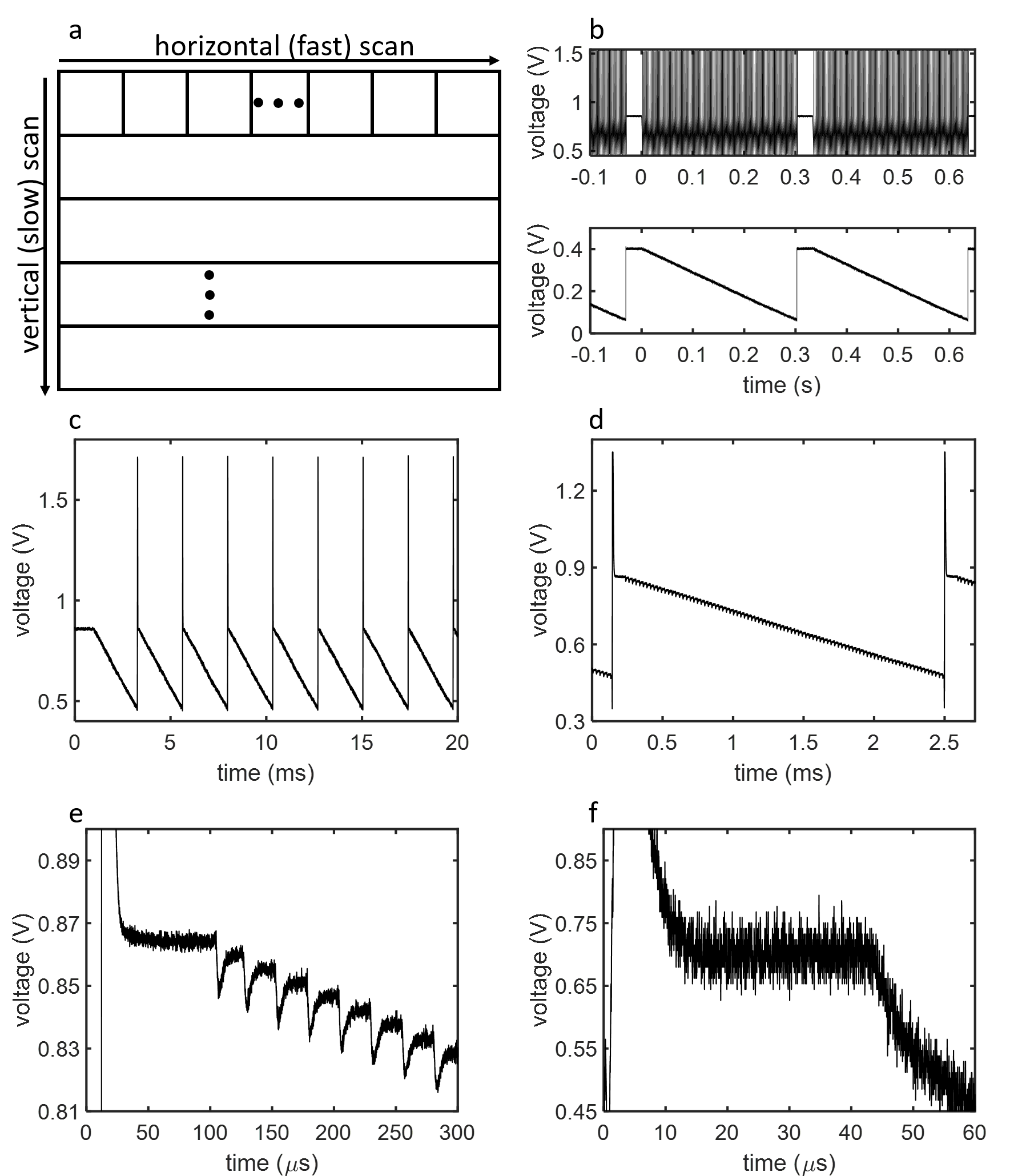}
    \caption[Scan waveforms on the SEM]{Scan waveforms on the SEM. (a) Schematic of SEM scanning, showing the two scan directions in an image frame: the fast horizontal scan and the slow vertical scan. (b) Horizontal (top) and vertical (bottom) scan waveforms for a pixel dwell time of 28 \us~and a pixel resolution of 129 lines by 88 pixels per line. Both the horizontal and the vertical scan waveforms are sawtooth-shaped. The total duration of the image frame in this example was 304 \ms. (c) First few lines in the horizontal scan. There is a 0.9 V spike at the end of each linescan, during the period where the electron beam is retraced to its horizontal starting position. (d) One linescan, showing the retrace spike, 70 \us~front porch, and trace region. The trace region has discrete steps corresponding to image pixels on the line. (e) Start of one linescan, showing the retrace spike, front porch, and trace with discrete steps. Before each step there is a transition region with a local minimum. The time duration between successive minima is 26 \us, close to the pixel dwell time. (f) Start of one linescan for a pixel dwell time of 440 \ns. Unlike (e), there are no discrete steps in this faster linescan.}
    \label{fig:figs1}
\end{figure*}

Figure~\ref{fig:figs1} (a) is a schematic of scanning in an SEM. The rectangle represents the area on the sample over which the SEM beam scans. An SEM image of this sample area consists of sequentially scanned lines (demarcated by horizontal bars in Figure~\ref{fig:figs1}(a)) with a certain number of image pixels on each line (indicated by the vertical bars within the first scan line in Figure~\ref{fig:figs1}(a)). There are two types of scan functions in all SEMs: the fast, horizontal scan and the slow, vertical scan. The horizontal scan defines one line in an image frame while the vertical scan defines one image frame. Here, an image frame refers to one scan of the incident electron beam over every pixel on the sample area being imaged. Therefore, within each image frame, the horizontal scan repeats for every line, while the vertical scan occurs once per image frame. 

Figure~\ref{fig:figs1}(b) is an example of the horizontal (top) and vertical (bottom) scan waveforms on the SEM. We obtained these waveforms for a scan area consisting of 129 lines by 88 pixels per line at a pixel dwell time of 28 \us. The figure contains the last few lines of  one image frame followed by two full image frames separated by a 30 \ms~frame reset time (the flat parts in the horizontal and vertical scans) . Both the horizontal and vertical scans consist of sawtooth waveforms. The horizontal scan has many repeated sawtooths in every image frame with each sawtooth waveform corresponding to one linescan. The resolution in the top image is too low to distinguish between the linescans. The vertical scan consists of one sawtooth waveform per image frame. The total duration of the image frame was 304 \ms~excluding the frame reset time at the start of every frame. 

In Figures~\ref{fig:figs1}(c) and (d) we take a more detailed look at the horizontal linescans of the first frame. Figure~\ref{fig:figs1}(c) shows the first few linescan waveforms, and Figure~\ref{fig:figs1}(d) shows one linescan. The total duration of each linescan is 2.45 \ms. In these and the following figures, we will see that each linescan consists of three regions, which we will call the retrace spike, front porch, and trace regions, following the terminology used for video scan signals. At the end of each linescan there is a 0.9 V voltage spike, lasting for 30 \us, before the scan waveform resets to the voltage at the start of the line. This resetting corresponds to the electron beam retracing to its horizontal starting position after every line. We can see these spikes at 0.1 and 2.5 \ms~in Figure~\ref{fig:figs1}(d). This voltage spike occurs because of ringing in the voltage of the scan electronics as it resets at the end of each line. Following the voltage spike there is a 70 \us~`plateau' (which we will call the front porch region) where the scan voltage is relatively flat. This region allows the scan voltage to settle to its starting value after the retrace spike and before the trace across the line on the sample begins. The front porch is followed by the trace region where the incident electron beam actually scans over the sample. This section of the scan waveform consists of a series of decreasing voltage steps, with each step corresponding to one pixel on the scan line. 

We can see these steps, along with the retrace spike and front porch regions of the linescan, in greater detail in Figure~\ref{fig:figs1}(e). Each step had a size of 4 mV and a time duration of 26 \us, measured at the valleys between each step. Since the number of pixels in each line was 88, we would expect the total voltage swing for one linescan to be 352 mV. This value is close to the observed swing of 380 mV in one linescan. Further, the time duration of each step was close to the pixel dwell time of 28 \us~calculated from the frame time specified for the scan speed chosen in the SEM software. Both these observations provide evidence for our assertion that these voltage steps corresponded to the individual pixels on each line of the image. We can also see that the steps are not completely flat; at the end of each pixel step there was a transition period before the scan waveform voltage level reached the value corresponding to the next pixel. This transition period was approximately 10 \us~long for the settings used in Figure~\ref{fig:figs1}(e) and was probably caused by the inherent speed of the scan coil electronics. Due to the presence of this transition period, we observed the discrete pixel steps for linescan waveforms corresponding to relatively long pixel dwell times only. For pixel dwell times below $\sim7.5$ \us, we did not observe this discretization and the trace decreased continuously. Figure~\ref{fig:figs1}(f) is an example of such a scan waveform at pixel dwell time of 440 \ns. This difference between slow and fast scans was important in the code we used to implement SE count imaging, as we discuss more extensively in Section~\ref{subsec:linepixel}.

\section{Extracting mean SE count rate from image histograms}

\begin{figure}
    \centering
    \includegraphics[scale = 0.48]{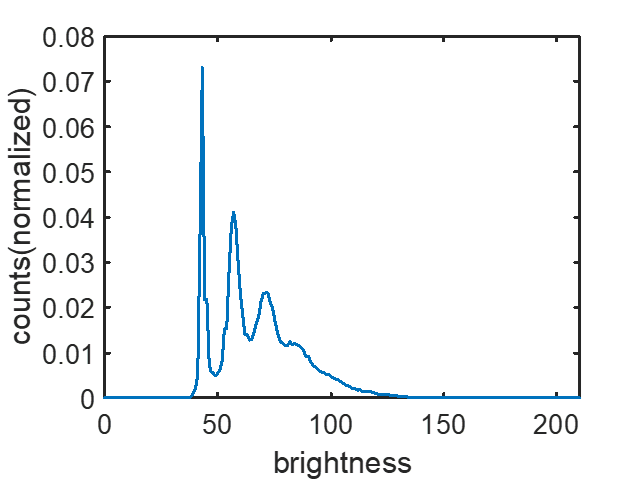}
    \caption{Image pixel brightness histogram for the same uniform aluminum sample used to characterize SE pulses in the paper. This histogram was acquired at a beam current of 2.3 $\pA$~and a pixel dwell time of 3.6 $\us$. The meant SE count extracted from this histogram is 1.45.}
    \label{fig:figs2}
\end{figure}

Figure \ref{fig:figs2} is an image pixel brightness histogram from the same aluminum sample scan region used to analyze SE detector pulses in Figure 2 in the paper, at a pixel dwell time of 3.6 $\us$ and a beam current of 2.3 pA. We have described such histograms in previous work. Each of the peaks in the histogram corresponds to pixels for which 0, 1, 2, and 3 SEs are detected. From this histogram, we found that the 1.45 SEs were detected from the sample on average. This average number of SEs corresponds to a mean SE detection rate of 0.41/$\us$, which is close to the theoretically expected value of 0.45/$\us$ calculated in the paper.

\section{Calculating image resolution from linescan waveforms}
\label{subsec:linepixel}
In our SE imaging code we used the linescan waveform to calculate the number of lines and the number of pixel per line in the image. Figure 3(a) in the paper shows two examples of the horizontal linescan voltage waveform. As can be seen in this figure, each line in the horizontal scan (the ramp portion of the waveform) was preceded by a retrace spike associated with the resetting of the scan voltage. Since one spike was associated with the end of one line, a count of the number of spikes was also a count of the number of lines in the image. Therefore, in the SE count imaging code, we used threshold discrimination to determine the number of spikes and hence the number of scan lines in the scanned sample area. 

We used a similar technique to calculate the number of pixels in every line. As discussed in Section~\ref{subsec:scanfeat} of the Supplementary Information, each pixel step in the trace section of a slow linescan was followed by a transition period during which the scan voltage transitioned to the next pixel. In Figure~\ref{fig:figs1}(e) we can see the scan voltage during this transition period had a minimum from which the voltage reached the next pixel voltage value in 10 \us. We used a count of these minima to count the number of pixels for each linescan. Discrete steps in the linescan corresponding to pixels were only present for slow linescans, and we did not observe them for linescans that corresponded to a pixel dwell time faster than 7.5 \us. This pixel dwell time was an order of magnitude larger than the pixel dwell time of 440 \ns~we used to generate the SE count images. Figure~\ref{fig:figs1}(f) shows the trace portion of the linescan sawtooth waveform for a pixel dwell time of 440 \ns, and we can see that it does not show discrete steps corresponding to each pixel in the linescan. Therefore, to count the number of pixels in the linescan, we acquired a slow linescan waveform (at pixel dwell time 28 \us) at the same image pixel resolution. Since we were only interested in collecting the scan waveform and not the SE signal from the object, we collected this slow linescan with the incident electron beam turned off to prevent unnecessary incident electron dose on the sample.

\section{Thong's method for finding signal-to-noise ratio (SNR) for an SEM image}
\label{sec:metric_SNR}
The SNR is a standard measure used in many fields of signal processing and imaging to characterize the quality of a signal. However, defining SNR for a single SEM image is challenging, because of the difficulty in differentiating signal from noise in an image of an unknown sample. A method for solving this problem was proposed by Thong et al. as referenced in the paper. They developed an SNR measure for a single SEM image by looking at the autocorrelation of the image. The autocorrelation function can be thought of as a measure of how similar an image is to an offset copy of itself. Here, we note that the autocorrelation is defined for a probabilistic model of a random process.  From a single realization of a random process, we cannot compute the autocorrelation, but we can compute an estimate of the autocorrelation of the underlying random process.  Here, we will model a noise-free image as a realization of a two-dimensional wide-sense stationary random process, and an observed signal is another two-dimensional wide-sense stationary random process with statistical dependence on the noise-free image. 

The autocorrelation $r_a(m)$ of a one-dimensional wide-sense stationary random signal $a(n)$ is defined as:
\begin{equation}
r_a(m) = \E\left[a(n)a(n-m)\right]
\label{eqn:autocorrdef}
\end{equation}
Here $m$ is called the offset or lag, and $n$ and $n-m$ are two points along the signal. $\E$ denotes expectation value. Note that the lack of dependence of the expectation on $n$ is a requirement of wide-sense stationarity. We would find the autocorrelation function of a physical image for a given pixel offset by multiplying it by a duplicate of itself offset by the required number of pixels. If the image has features that extend over a many pixels, a displaced copy of the image will have similar intensities in the same pixels as the original image. Therefore, the autocorrelation of such an image would be high. Conversely, if the pixel values in the image vary rapidly over a few pixels, the autocorrelation will be small.

Here, we will describe Thong et al.'s SNR measure and present a new, rigorous justification behind the measure. This SNR measure is based on the assumption that the contribution to the image autocorrelation from the `signal' component of the image varies slowly (\textit{i.e.}, over several pixels), whereas the `noise' component of the autocorrelation varies rapidly and is zero for $m\neq0$ in the definition of autocorrelation in Equation~\ref{eqn:autocorrdef}. This difference in the nature of signal and noise allows the extraction of their relative contributions to the autocorrelation, and the ratio of these contributions is a measure of the image SNR.

We will first consider an ideal Poisson signal derived from an underlying ground truth image and analyze its autocorrelation. Then, we will introduce some of the non-idealities present in an SEM and derive Thong et al.'s expression for the SNR measure. Finally, we will demonstrate the effectiveness of the SNR measure by calculating it for SEM images taken at different pixel dwell times.

Let $f(n)\sim\textrm{Poisson}(s(n))$ be a Poisson process derived from $s(n)$. We assume that if $s(n)$ is known, the values of $f(n)$ at two points $n_1$ and $n_2, n_1\neq n_2,$ would be independent of each other. This assumption is equivalent to assuming that the noise for any pair of pixels is uncorrelated (and hence has an expected value of zero). We want to find the autocorrelation of $f(n)$, \textit{i.e.}, we want to find $r_f(m)=\E\left[f(n)f(n-m)\right]$. We consider two cases: $m\neq0$ and $m=0$.

\textbf{Case 1: $m\neq0$}
\begin{flalign*}
 & \E\left[f(n)f(n-m)\right] \\ 
 & \ =\ \E\left[\E\left[f(n)f(n-m)|\lbrace s(n) \rbrace\right]\right] \\
 & \ =\ \E\left[\E\left[f(n)|{s(n)}\right]\E\left[f(n-m)|\lbrace s(n)\rbrace\right]\right] \\
 & \ =\ \E\left[s(n)s(n-m)\right] \\
 & \ =\ r_s(m)
\end{flalign*}
Here, the second equality results from our assumption about the independence of $f(n_1)$ and $f(n_2)$, and the third equality results from the definition of $f(n)$. Therefore, we conclude that $r_f(m)=r_s(m)$ for $m \neq 0$.

\textbf{Case 2: $m=0$}
\begin{flalign*}
\E\left[f(n)f(n-m)\right] & \ =\ \E\left[f(n)^2\right] \\
 & \ =\ \E\left[\E\left[f(n)^2|\lbrace s(n)\rbrace\right]\right] \\
 & \ =\ \E\left[s(n)^2+s(n)\right] \\
 & \ =\ r_s(0) + \E\left[s(n)\right]
\end{flalign*}
Here, the third equality again results from the definition of $f(n)$. Therefore, we finally get:
\begin{equation*}
    r_f(m) \ = \ \left\{\begin{array}{@{\ }ll}
 r_s(m), & m\neq0; \\
 r_s(0)+\E\left[s(n)\right], & m=0.
 \end{array}\right.
\end{equation*}
Hence, the autocorrelation $r_f(m)$ of the noisy image $f(n)$ derived from the ground truth $s(n)$ will be the same as the autocorrelation of $s(n)$ everywhere except at $m=0$, where there is an additional term due to the fact that $f(n)$ is derived from a random process $s(n)$ and is therefore noisy. As we had stated earlier, the assumption we made about the independence of values of $f(n)$ at different pixels is equivalent to assuming that the noise is uncorrelated between neighbouring pixels, which is why the contribution due to noise only affects the autocorrelation at zero offset. At zero offset, we expect the autocorrelation to have a sharp peak due to the additional noise contribution. At non-zero offsets, this noise contribution is zero because the noise in neighbouring pixels is uncorrelated.

Our assumption that $f(n)\sim\textrm{Poisson}(s(n))$ is not justified for conventional SEM imaging, since a conventional SEM image is not a count of SEs per pixel. Further, in addition to noise due to the randomness inherent in the generation of SEs, the SE image can also have noise due to imperfect detection of SEs, background counts in the detector, variation in the signal waveform from the detector, and quantization of the SE signal. We can model the scaling to 8-bit and noise addition in an SEM as a linear scaling of the number of SEs observed for that pixel, along with zero mean Gaussian noise, as follows:
\begin{equation*}
    f(n)\sim \textrm{A}\cdot\textrm{Poisson}(s(n))+N(0,\sigma^2)
\end{equation*}
Here, the factor A accounts for scaling to 8 bits, and $N(0,\sigma^2)$ accounts for the additional sources of noise discussed above. Using the same process to find the autocorrelation $r_f(m)$, we get:

\begin{flalign}
     & r_f(m) \\
     & \ = \ \left\{\begin{array}{@{\ }ll}
 \textrm{A}^2 r_s(m), & m\neq0; \\
 \textrm{A}^2 r_s(0)+\textrm{A}\E\left[s(n)\right]+\sigma^2, & m=0.
 \end{array}\right.
 \label{eqn:autocorr}
\end{flalign}
Hence, for a real image, we would still expect the autocorrelation to show a sharp peak at zero offset. Now, the peak at zero offset has contributions from both noise due to the Poisson process and the variance of the additional Gaussian noise. Thong's method provides a way to estimate the ``noise-free'' component of the autocorrelation, at zero offset ($\hat{\phi}_{NF}(0,0)$, an estimator for the true noise-free autocorrelation $\phi_{NF}(0,0)=\textrm{A}^2 r_s(0)$) from the observed autocorrelation at zero offset $\phi(0,0)$) and estimate both the signal and noise contributions in the image to derive SNR for a single SEM image. Note that our SEM images are two-dimensional and hence we specify the offset along both dimensions in our notation. We will demonstrate this method through an example.

\begin{figure*}
    \centering
    \includegraphics[scale=0.48]{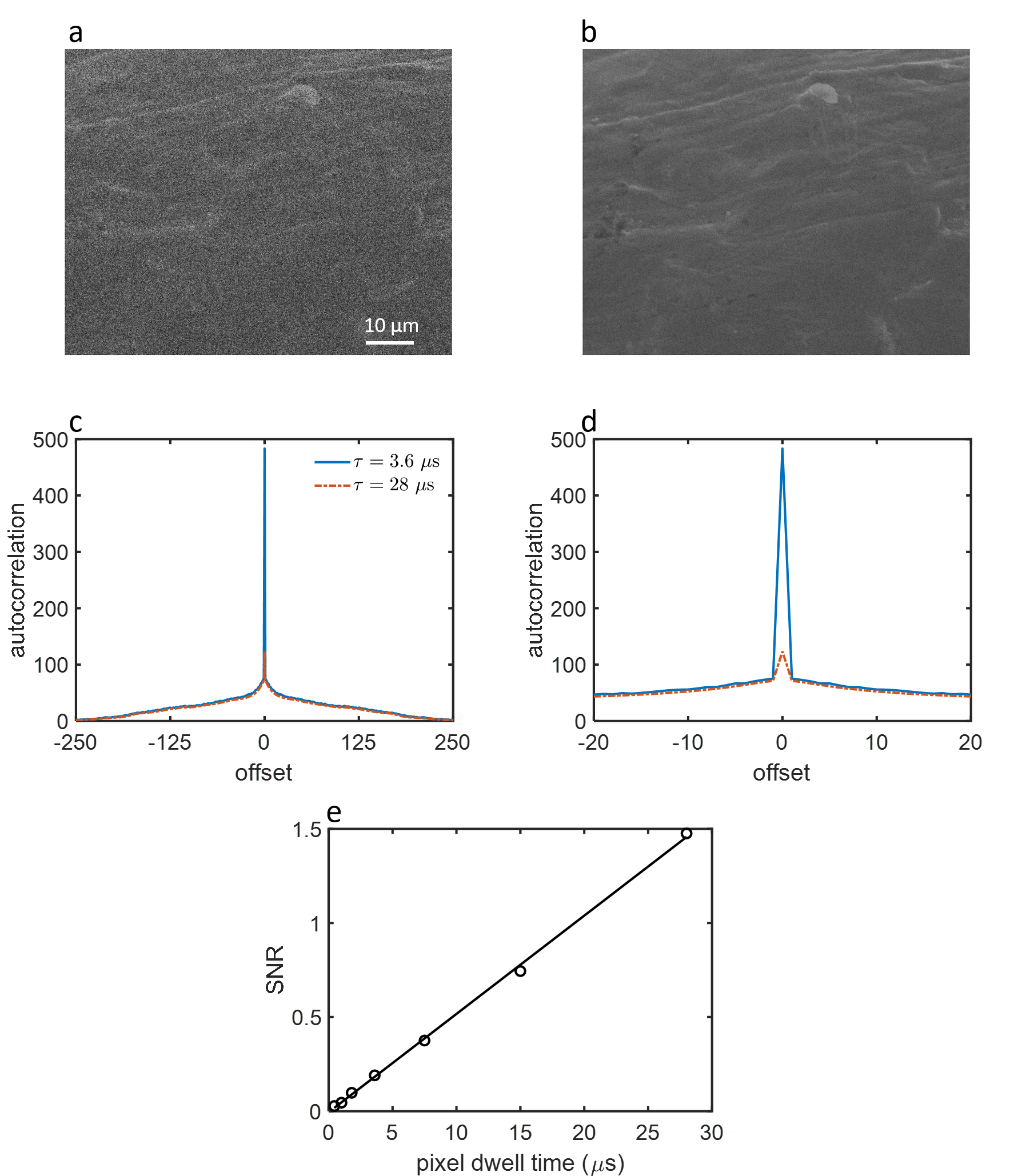}
    \caption[SNR for SEM images]{SNR for SEM images. (a) SEM image of bulk copper, taken at a pixel dwell time of 3.6 \us. (b) SEM image of the same sample as (a), taken at a pixel dwell time of 28 \us. (c) Autocorrelation of the 3.6 \us~dwell time image (solid blue curve) and the 28 \us dwell time image (dash-dotted orange curve). The autocorrelation is almost identical except at zero pixel offset. (d) Autocorrelation of the two images around zero pixel offset. The sharp peak at zero offset is because of image noise. The peak is lower for the 28 \us~pixel dwell time image because it is less noisy. (d) SNR extracted from the autocorrelation of SEM images acquired with pixel dwell times between 0.44 \us~and 28 \us. The SNR scales linearly with pixel dwell time.}
    \label{fig:figs3}
\end{figure*}

Figures~\ref{fig:figs3}(a) and (b) are two SEM images of the same region of a bulk copper sample, taken at the same incident beam current. We acquired Figure~\ref{fig:figs3}(a) at a pixel dwell time of $3.6$ \us~and Figure~\ref{fig:figs3}(b) at a pixel dwell time of $28$ \us. The longer dwell time we used to acquire Figure~\ref{fig:figs3}(b) resulted in a less noisy image.

In Figure~\ref{fig:figs3}(c), we plot the autocorrelations of these two images. The full autocorrelation is a three-dimensional function since the images can be offset in two dimensions. In Figure~\ref{fig:figs3}(c) we have plotted the autocorrelation for offsets in the horizontal direction; the offset in the vertical direction is zero. The solid blue curve is the autocorrelation of the image in Figure~\ref{fig:figs3}(a), and the dash-dotted orange curve is the autocorrelation of the image in Figure~\ref{fig:figs3}(b). We can see that the autocorrelation values for the two images are almost identical at all offsets except 0. Figure~\ref{fig:figs3}(d) is a magnified view of the autocorrelations around zero offset. We can see that the less noisy, longer pixel dwell time image has a smaller autocorrelation peak at zero offset compared to the higher noise, shorter pixel dwell time image.

From Figure~\ref{fig:figs3}(d) we can also see that the autocorrelation curves for small, non-zero offsets are pretty flat, especially when compared to the sharp peaks at zero offset. For example, between pixel offset values of 1 and 10, the autocorrelation of the $\tau=3.6$ \us~image reduces from 75 to 46.8 (an average reduction of 3.1 per additional pixel offset), whereas it jumps to 483.4 at zero offset (an increase of 408.4 from the value at pixel offset of 1). We can use this observation to estimate the noise-free autocorrelation at zero offset, $\hat{\phi}_{NF}(0,0)$. Thong discusses two methods of finding $\hat{\phi}_{NF}(0,0)$. The first method is to simply set it to be equal to the value of the autocorrelation at an offset of 1 pixel, \textit{i.e.}, $\hat{\phi}_{NF}(0,0)=\phi(1,0)$. The second method of estimating $\hat{\phi}_{NF}(0,0)$ is to use linear extrapolation of the values of the autocorrelation for small pixel offsets. Both these methods are justified because, as we had discussed, the autocorrelation value does not change significantly between consecutive pixels. We decided to use the second method to extract $\hat{\phi}_{NF}(0,0)$ because, as we can see in Figure~\ref{fig:figs3}(d), the autocorrelation is close to linear at small pixel offsets. 

Once we have extracted $\hat{\phi}_{NF}(0,0)$, we can obtain an estimate of the noise contribution $\hat{\phi}_{N}(0,0)$ to $\phi(0,0)$: $\hat{\phi}_{N}(0,0)=\phi(0,0)-\hat{\phi}_{NF}(0,0)$. The ratio of these two quantities gives us a measure of the image SNR:
\begin{equation*}
    \textrm{SNR}=\frac{\hat{\phi}_{NF}(0,0)}{\hat{\phi}_{N}(0,0)}
\end{equation*}

We used this method to find the SNR for images of the same sample of bulk copper as in Figure~\ref{fig:figs3} for pixel dwell time between 0.44 \us~and 28 \us.

Figure~\ref{fig:figs3}(e) is a plot of the extracted SNR values, indicated by unfilled black circles. The solid black line is a least-squares linear fit to the extracted values. We can see that the extracted SNR values scale almost linearly (correlation coefficient $r^2>0.99$) with the pixel dwell time. We can derive this linear scaling from our expression for $r_f(m)$. Due to the fact that the conventional SEM image pixel intensity is the average value of the detector signal observed for that pixel, increasing the pixel dwell time does not change the mean value of the pixel intensities: although the total signal observed for each pixel increases, the pixel dwell time increases by the same factor. Therefore, the mean pixel value remains the same. We can include this effect in our model of a real SEM image as follows: let the two pixel times $t_1$ and $t_2$ correspond to two images $f_1$ and $f_2$. Let $t_2/t_1=\alpha$. On changing the dwell time from $t_1$ to $t_2$, the raw signal from the SE detector will scale by $\alpha$. Therefore, we have:
\begin{flalign*}
     & f_1(n)\sim \textrm{A}_1\cdot\textrm{Poisson}(s(n))+N(0,\sigma) \\
     & f_2(n)\sim \textrm{A}_2\cdot\textrm{Poisson}(\alpha s(n))+\alpha N(0,\sigma)
\end{flalign*}
Due  to signal averaging, $\E\left[ f_1(n)\right]=\E\left[ f_2(n)\right]$. Hence, $\textrm{A}_1/\textrm{A}_2=\alpha$. Using Equation~\eqref{eqn:autocorr} to find the autocorrelation $r_f(0)$, we get:
\begin{flalign*}
 & r_{f_1}(0)=\textrm{A}_1^2 r_s(0)+\textrm{A}_1^2\E\left[s(n)\right]+\sigma \\
 & r_{f_2}(0)=\alpha^2\textrm{A}_2^2 r_s(0)+\alpha\textrm{A}_2^2\E\left[s(n)\right]+\alpha^2\sigma \\
\end{flalign*}
Here, the expression for $r_{f_2}(0)$ results from the fact that scaling a random variable by $k$ scales its variance by $k^2$. Simplifying this expression for $r_{f_2}(0)$, we get:
\begin{equation*}
r_{f_2}(0)=\textrm{A}_1^2 r_s(0)+\textrm{A}_1^2\E\left[s(n)\right]/\alpha+\alpha^2\sigma
\end{equation*}

Comparing this expression to the expression for $r_{f_2}(0)$, we can see that the noise-free component of the autocorrelation is the same for both but the contribution due to noise due to the Poisson nature of the signal has scaled down by $\alpha$. The noise from other sources in the SEM scales by $\alpha^2$. For the imaging conditions normally used in an SEM, the contribution of noise from other sources on the SEM is small and the $\sigma^2$ term can be ignored. Consequently, the SNR scales linearly with the pixel dwell time, just as we had observed in Figure~\ref{fig:figs3}(e).

In summary, this SNR measure provides another way to characterize image quality and scales linearly with the incident electron dose. We require two conditions to be fulfilled for this method to work: the noise autocorrelation must be non-zero only at a pixel offset of 0, and the autocorrelation of the ground truth image must vary much more slowly than the noise autocorrelation. From the autocorrelation curves in Figure~\ref{fig:figs3}(c) and (d) we can see that noise in the SEM images does not show correlation for any non-zero pixel offsets. We also confirmed that this was the case by scanning the beam over a region of vacuum and checking the autocorrelation of such images. Further, we ensured that all samples we scanned had features that extended over many pixels so that the image autocorrelation varied slowly at non-zero pixel offsets. We note that the absolute value of this SNR measure for a particular SEM image is not very informative. For example, the SNR value for the long-dwell time image in Figure~\ref{fig:figs3}(b) is about 1.5, which would be considered low in other scenarios. It is the change in the value of this SNR metric upon varying the imaging conditions which provides useful information about changes in the image quality. 

\section{Probability distribution of SEs}
\label{subsec:SEdist}
\begin{figure*}
    \centering
    \includegraphics[scale=0.48]{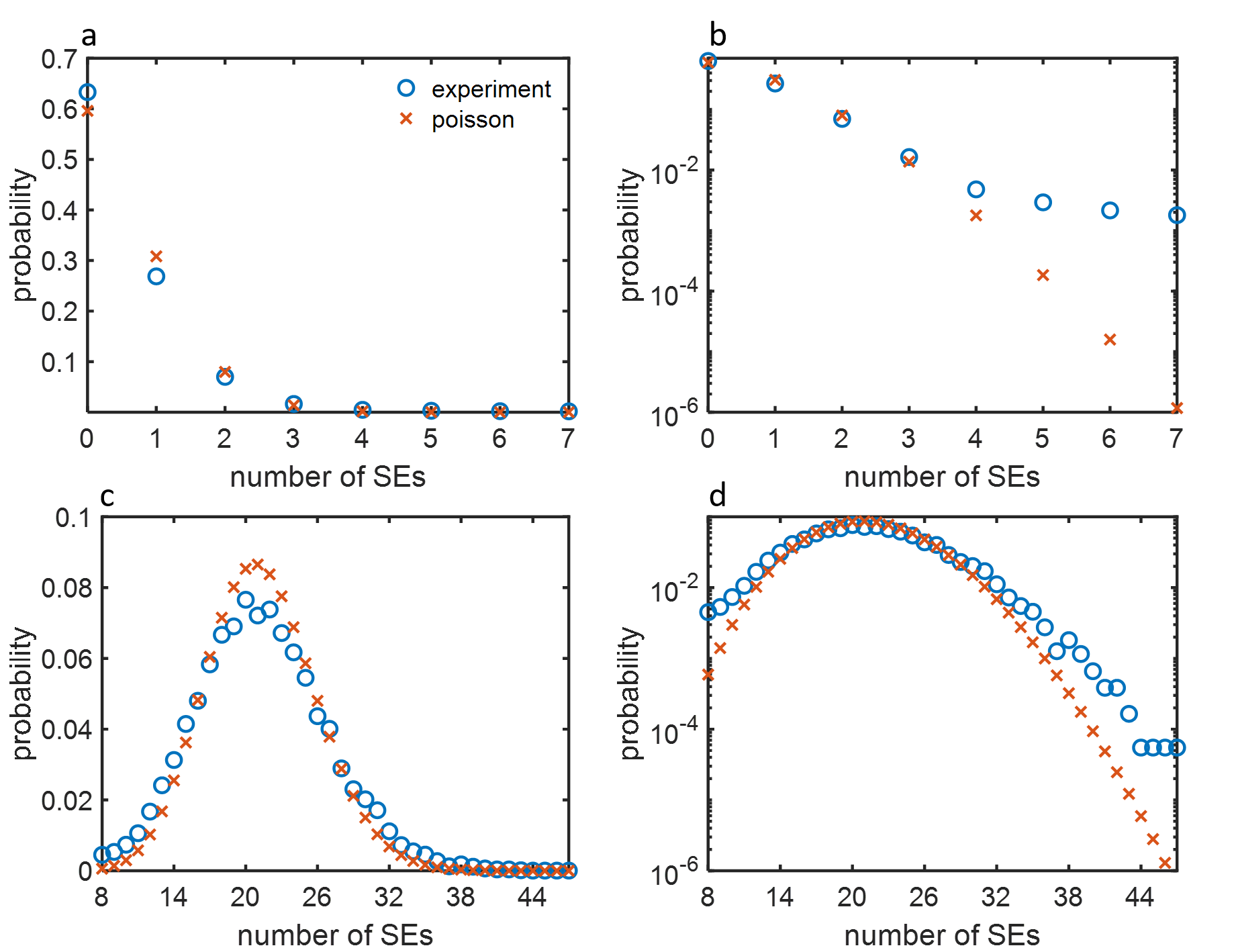}
    \caption[Probability distribution of SE emission]{Probability distribution of SE emission. (a) Distribution of the number of SEs extracted from the background (vacuum) pixels (unfilled blue circles). The mean SE count for this distribution is 0.52 and the variance was 0.74. The orange crosses indicate the probabilities of a Poisson process with the same mean SE count. (b) Same as (a) with the y-axis on a log scale. The extracted SE distribution deviates significantly from an ideal Poisson distribution beyond 3 SEs, reflecting its larger variance. (c) Distribution of the number of SEs extracted from the sample (copper mesh) pixels (unfilled blue circles). The mean SE count for this distribution is 21.29 and the variance was 29.8. The orange crosses indicate the probabilities of a Poisson process with the same mean SE count. The Poisson distribution has a higher and narrower peak than the extracted SE distribution. (d) Same as (c) with the y-axis on a log scale. The extracted SE distribution deviates significantly from the ideal Poisson distribution above 30 SEs and below 14 SEs, reflecting its larger variance.}
    \label{fig:figs4}
\end{figure*}

In this section, we will show a preliminary analysis of the probability distribution of the number of SEs in our experiments. The sample we imaged contained two types of pixels - bright `sample' pixels that represent the copper mesh, and dark `background' pixels that represent the background vacuum. Therefore, we found the statistics of these two types of pixels separately.

Figure~\ref{fig:figs4}(a) is a plot of the relative count of pixels that represent the background vacuum (unfilled blue circles) as a function of the number of SEs in those pixels. In this plot, we have normalized the counts so that they sum to 1 over the different numbers of SEs. Therefore, this plot can be thought of as a probability distribution of the number of SEs. The mean SE counts for these background pixels was 0.52. To visualize the deviation from an ideal Poisson process, we have also plotted the probabilities for a Poisson process with the same mean (orange crosses). We can see that the probability of 0 SEs is higher and that of 1 SE is lower for the distribution we extracted from our data than for the ideal distribution. Figure~\ref{fig:figs4}(b), which is the same as \ref{fig:figs4}(a) but with the y-axis on a log scale, shows that at higher SE numbers the extracted probabilities are much higher than the probabilities from the Poisson distribution. From these plots, we would expect the variance of the extracted probabilities to be higher than the variance for an ideal Poisson process, which should be equal to 0.52. The variance of the extracted distribution is 0.74, confirming this intuition. From these numbers, the Frank's B-factor of this process is 0.42. 

We performed a similar calculation for the pixels representing the copper mesh. Figure~\ref{fig:figs4}(c) is a plot of the  relative number of copper mesh pixels (unfilled blue circles) as a function of the number of SEs in those pixels. Figure~\ref{fig:figs4}(d) is the same plot with the y-axis on a log scale. Just as in Figure~\ref{fig:figs4}(a) and (b), we scaled the pixel numbers so that they summed to 1. we have also plotted the probabilities of an ideal Poisson process with the same mean as the extracted distribution (21.29) using orange crosses. From these plots we can see that the extracted distribution was wider than an ideal Poisson distribution. The variance of the extracted distribution was 29.8. This variance value gave us a B-factor of 0.4, which is very close to the B-factor we had extracted from the vacuum pixel distribution. We note that this B-factor is about 2 times larger than that reported by Frank for copper at 10 keV. We expect this deviation to be caused by small non-uniformities in the sample, as well as the enhancement of SE yield at the edges of the copper grid. Mitigation of these effects by imaging a uniform sample without edges would provide more accurate SE probability distributions using our technique.

\section{Schemes for online electron counting and conditional re-illumination}
\label{sec:online}
 
As discussed in the paper, we cannot use the SE count imaging scheme discussed here to realize the potential dose reduction due to counting because the scanning of the beam over the sample was not turned off automatically between acquisition frames. Saving each acquisition frame on the oscilloscope took a few seconds, and the beam was scanning over the sample for some part of this time before we turned it off manually. Therefore, the sample was exposed to the electron beam between acquisition frame collections on the oscilloscope. This additional electron dose offset any reduction in dose we might have obtained with SE count imaging.

In a live implementation of SE count imaging, we would want each imaging frame (corresponding to the beam scanning over the sample) on the SEM to correspond to an acquisition frame in the dataset. Therefore, the ideal workflow would be to blank the beam while an acquisition frame is being recorded by the oscilloscope, and unblank the beam when the oscilloscope is ready to receive the next frame. This blanking would need to be synchronized with the SEM scan waveforms so that each of the acquisition frames begins with the first line of the scanned area. In this section, we will discuss such a scheme to implement live, online SE counting.

\begin{figure*}
    \centering
    \includegraphics[scale=0.44]{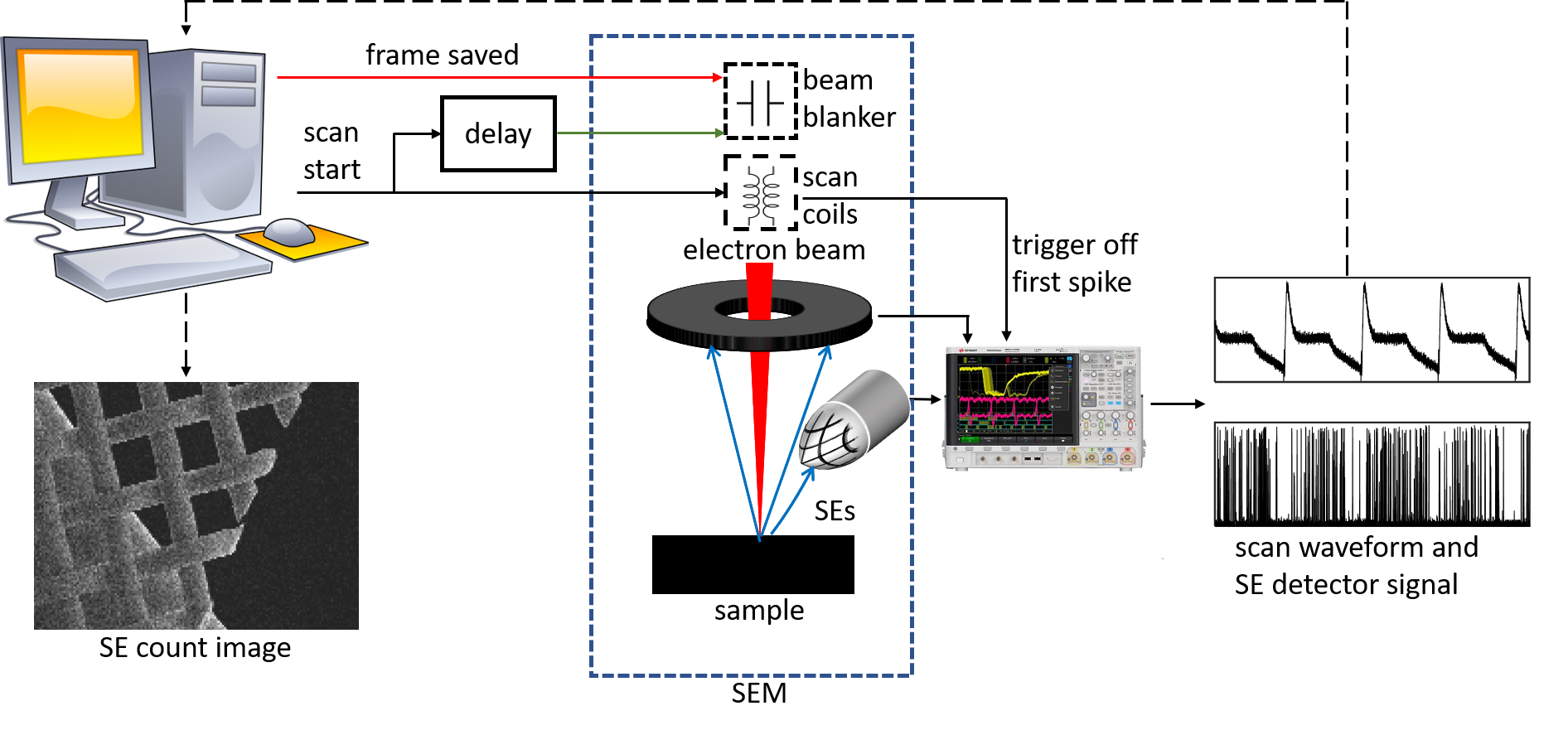}
    \caption[Scheme for live SE count imaging]{Scheme for live SE count imaging. Solid black arrows indicate online steps, and dashed black arrows indicate offline steps. The scan start signal from the SEM computer, with a delay equal to the acquisition frame time, activates the beam blanker (green arrow). After the frame is acquired on the oscilloscope, the SEM computer sends an unblanking signal (red arrow) to the beam blanker before the start of the next frame. After all frames have been acquired, the SE count image can be generated on the SEM computer offline.}
    \label{fig:figs5}
\end{figure*}

Our scheme for live SE count imaging is shown in Figure~\ref{fig:figs5}. At the start of the imaging process, the SEM computer sends a signal to the SEM scan coils to start the first frame. The oscilloscope captures the scan waveform and the synchronized SE detector signal pulses. The only addition in this scheme compared to the one we have implemented in the paper would be a beam blanking signal sent from the SEM scan computer to the SEM beam blanker after every acquisition frame. From our study of the scan waveform as discussed in Section~\ref{subsec:scanfeat}, we know the total acquisition frame time and the frame reset time for a given image resolution. Therefore, the initial signal from the SEM computer, with an added delay equal to the acquisition frame time, can serve as the blanking signal the beam blanker. This signal is shown by the green arrow in Figure~\ref{fig:figs5}. This delay could be introduced using a simple delay line generator or a comparator circuit. On receiving this signal, the beam blanker blanks the SEM before the linescans for the next frame can begin, thereby protecting the sample from additional electron dose while the previous acquisition frame is being saved. As we had described in the paper, noise in the scan signal can lead to variation in which voltage spike triggers data collection on the oscilloscope. With the trigger voltage set precisely, we observed that this uncertainty was usually between 1-3 lines, and at most 10 lines, which corresponds to a time delay of $\sim1$ \ms. Since the frame reset time is around 30 \ms~depending on the pixel dwell time and scan area size, we have sufficient time to send the blanking signal even with possible delays due to noise. The time taken to save an acquisition frame on our oscilloscope was constant for a fixed sampling time and frame size. For example, for the frame parameters in the paper, the time to save was around 5 s. Since we know when scanning began and the frame times, the SEM computer already has an internal clock synchronized with the scanning frames. Therefore, after the time to save the previous frame has elapsed, the SEM computer sends an unblanking `frame saved' signal (shown by the red arrow in Figure~\ref{fig:figs5}) to the beam blanker before the start of the next scan frame. The process repeats for capturing the desired number of acquisition frames. After all the acquisition frames have been collected, the frames are transferred to the SEM computer and the computer processes the frames to generate the final SE count image. These offline processes are indicated with dashed arrows in Figure~\ref{fig:figs5}. Alternatively, they can be brought online and the signal to unblank the beam (indicated in red in Figure~\ref{fig:figs5}) can be sent after the acquisition frames are transferred to the computer and the count image is generated.

The major advantage of this scheme over Yamada's live SE counting scheme is that we have captured all the fast timing and processing complexity in the SE count imaging code instead of needing to implement it in hardware circuits as in Yamada's work. The only hardware timing signal is the one sent to the beam blanker after every acquisition frame. Since each frame typically lasts several tens to hundreds of \ms~depending on the image size, we do not need any high-speed electronics to implement this signal. As we discussed in the previous paragraph, we need to blank the beam during the frame reset time of 30 \ms~after the completion of one imaging frame before the next one can begin. The electrostatic beam blanker already built-in the SEM and used for freezing the SEM scan is sufficiently fast to blank the beam during this time. The resonant frequency of the build-in blanker on our SEM was around 10 \si{\kilo\hertz}, and the blanker did not blank the beam effectively above 20  \si{\kilo\hertz}. This speed corresponds to a blanking time of about 50 \us, which is much lesser than the frame reset time. Therefore, such a blanker would be good enough for this implementation of live SE count imaging.

\bibliography{references}
\bibliographystyle{unsrt}